\documentclass[english,british,preprint,aps,12pt,showpacs]{revtex4}
\usepackage{graphicx}
\usepackage{babel}
\usepackage{bm}
\usepackage[T1]{fontenc}

\def\s{\sigma}

\begin{document}
\preprint{}
\title{Thermodynamic properties of the XX model in a chain with a period two and three coupling}
\author{Julio Abad}
\affiliation{
Departamento de F\'{\i}sica Te\'orica,
Facultad de Ciencias,
Universidad de
Zaragoza, 50009 Zaragoza, Spain.}

\date{\today}

\begin{abstract}
The exact solutions for the energy spectrum of the XX model with a periodic
coupling and an external transverse magnetic field $h$ are obtained.  The
diagonalization procedure is discussed, and analytical and numerical solutions
are given.  Using the solutions for period-two coupling, the free energy,
entropy, and specific heat are calculated as functions of temperature and
applied transverse external magnetic field.  Their expressions show that below
a particular value $v$ and above a value $u$ of the magnetic field $|h|$, the
entropy and the specific heat vanish exponentially in the low temperature
limit.
\end{abstract}


\maketitle

\section{\label{intro}Introduction}

The physics of one dimensional chains provide a good example for the rigourous
study of some properties of quantum magnetic systems.  They are known to be
useful for analysing many-body problems.  Recently, real, model quasi-1D
systems of atoms have become the object of experiments.  From a experimental
point of view, there are many quasi one dimensional compounds (such as the
organic series (BCPTTF)$_{2}X$ with ($X=$ AsF$_{6}$, PF$_{6}$), the cuprate
CuGeO$_{3}$, or the (VO)$_{2}$P$_{2}$O$_{7}$ compounds, as well as other series
as the TTFMS$_{4}$C$_{4}$(CF$_{3}$)$_4$ with $M=$ Cu, Au, Pt, or Ni
\cite{bra75,has193,has293,bou96,joh87,nig86,whi93}).  These are properly
described as Peierls spin-chains.

The experimental discovery of quasi-crystals~\cite{sch,lev} has stimulated
intense activity in the study of the periodic and aperiodic chains, with the
aim of understanding their physical properties~\cite{luc1}.

Recently, new, unusual properties in low-dimensional magnetic materials have
been found~\cite{dag96,ngu96,gam02} that can be explained in terms of many-body
behaviour.  It seems to be responsible for the occurrence of magnetization
plateaus as a function of external magnetic fields~\cite{bos03,oka02}.

The procedure for the study of this type of structures was initiated by Leib
\textit{et al.}~\cite{lieb}, by using the well known Jordan-Wigner
transformation~\cite{wig1}, and assuming that the chain is finite with set
boundary conditions.

One of the most important systems is the 1D $XY$ model ($S=1/2$), introduced by
Leib~\textit{et al.}~\cite{lieb}.  It plays an important role in the
description of many-body problems since it can be solved exactly.


Two early examples of studies of ground state properties (\textit{e.g.}
susceptibility) for the $XY$ model with alternating interaction are those by
Perk \textit{et al.}~\cite{per}, and Taylor and M\"{u}ller~\cite{tay}.

Quantum critical phenomena in random $XY$ chains have been studied using
renormalization group methods~\cite{fis}, and by numerical
methods~\cite{luc2,you}.

Recently, Derzhko \textit{et al.}~\cite{der1,der99,der2} have described the
thermodynamical behaviour and properties of the ground state of periodic
non-uniform $XY$ chains in a transverse field.  Also, thermodynamic properties
in one dimensional superlattices have been treated by de Lima \textit{et
al.}~\cite{lim95,lim1,lim99}.  An excellent review is provided in
Ref.~\cite{lim2}.

As mentioned previously, systems of this type are interesting due to being
strongly affected by quantum fluctuations, and by their apparent simplicity.
For example, low dimensional electronic materials are known to be very
sensitive to structural distortions that are driven by electron-phonon
interactions.  These break the symmetry of the original ground state, which
results in a new lower energy state where the electrons and ions are shifted
from their original positions in a regular manner.  This creates a periodic
variation of the charge density that is called a charge density wave.  This is
the well know Peierls instability which opens a gap at the Fermi surface of the
1D electronic chain, transforming a metal into a semiconductor.  A similar
effect is expected for 1D spin chains, which are unstable against lattice
vibrations.

The problem considered here is a homogeneous chain where the structural
distortions of the lattice are taken into account through a periodic coupling
between a given site and its neighbouring sites.
An external transverse magnetic field $h$
is also applied.  This permits us to study the diverse phases of the system.
Thus, the Hamiltonian is,

\begin{equation}
H\,=\,-\sum_{i=1}^N J_{i,i+1} (S^x_i S^x_{i+1}+ S^y_i S^y_{i+1}) -
h \sum_{i=1}^N S^z_i,\label{heis}
\end{equation}
where
\begin{equation}
J_{i,i+1}\,=\,J_{i+a,i+a+1}\,=\,u + v\,\cos\left(\frac{2\pi m}{N}i\right),
\label{heis2}
\end{equation}
and the quantum spin operators are $S_i^c = (1/2)\, \sigma_i^c$ (where
$\sigma_i^c$ are the Pauli matrices at the site $i$), and $a\,=\,N/m$ is an
integer.

This work is divided in four sections.  In section~\ref{method} we begin with
the Hamiltonian (\ref{heis}).  The Jordan-Wigner transformation is applied,
then the Hamiltonian is diagonalized with a superposition between creation
operators.  The same superposition is used between the corresponding
destruction operators.  This superposition is possible due to the Hamiltonian
being isotropic, thus preserving the total transverse spin $S_z=\sum_i S_i^z$.

In section~\ref{results}, analytical results for the energy bands and gaps are
given for the integer values $a=2$ and $3$.  Graphs are provided that show
numerical results for higher values of $a$.  The degeneration of the bands into
single levels is discussed.

In section~\ref{thermo}, thermodynamic quantities---free energy, entropy and
specific heat---and their behaviour are discussed.  The free energy per site
is calculated as a function of temperature.  The other thermodynamic properties
are subsequently derived from this.  Graphs showing the free energy and entropy
are given.

An interesting result is found by taking the low temperature limit.  It is
shown that the entropy and the specific heat possess local maxima at $|h|=v$
and $|h|=u$.  They vanish exponentially with temperature for $|h|<v$ and
$|h|>u$.  Both functions at the maxima vary as a square root of temperature,
while in between they are proportional to temperature.

\section{\label{method}The model}

The simplest solvable model (see for example Ref.~\cite{ta1}) is the isotropic
spin $1/2$ $XY$ model.  This model is based on a one dimensional lattice with a
uniform interaction strength $J$ between nearest neighbour sites, and uses
periodic boundary conditions.  As explained in the introduction, a correction
to this Hamiltonian is provided by introducing a periodic strength, which
simulates a vibration at the sites.  Hence, the Hamiltonian from (\ref{heis},
\ref{heis2}) is
\begin{equation}
H\,=\,-\sum_{i=1}^N \left[u + v \cos\left(\frac{2 \pi m}{N}i\right)\right]
\left(S^x_i S^x_{i+1}+S^y_i S^y_{i+1}\right)-h S^z_i,\label{hxx}
\end{equation}
where $u$ is a uniform coupling and $v$ is the amplitude of the periodic
coupling.

The spin operators $S_i$ act on a two dimensional space $\eta_i= C^2$ where we
take the eigenstates of $S^3$ as a base.  The total third component of the
spin,
\begin{equation}
S_z\,=\,\sum_{i=1}^N S_i^z,
\end{equation}
is a conserved quantity that can be used to describe the states of the system.

The first task is to transform the Hamiltonian (\ref{hxx}) into a fermionic
Hamiltonian by means of the Jordan-Wigner transformation \cite{wig1}.  This is
achieved by first defining the fermionic operators:
\begin{eqnarray}
a_l&=&K(l)\, S^{-}_l\,,\\
a^{\dagger}_l&=& K(l)\,S^{+}_l, \label{v}
\end{eqnarray}
where $S^{\pm}_l=\left(S^x_l \pm i S^y_l\right)$, and
\begin{equation}
K\left(l\right)\,=\,\exp\left(i \pi \sum_{j=1}^{l-1}
S^{+}_j S^{-}_j\right)\,=\,\prod_{j=1}^{l-1}\left(-2 S^z_j\right).
\label{vi}
\end{equation}

The new Hamiltonian and the conserved spin component $S_z$ become,
\begin{equation}
H\,=\,-\sum_{n=1}^N \left\{\left[u+v\cos\left(\frac{2 \pi}{a}n\right)\right]\,
\frac{1}{2}\,\left(a^{\dag}_n a_{n+1}+a^{\dag}_{n+1}
a_{n}\right)- h\left(a^{\dag}_{n}a_{n}-\frac{1}{2} I\right)\right\}
\label{vii}
\end{equation}
\begin{equation}
S_z\,=\, \sum_{n=1}^N\left(a^{\dag}_{n}a_{n}-\frac{1}{2} I\right)\,=\,
N_f\,-\,\frac{N}{2}I
\end{equation}
$I$ is the unity operator, and $N_f=\sum_{n=1}^N\left(a^{\dag}_{n}a_{n}\right)$
is the total fermionic number, which is also conserved.

These $a$ and $a^\dagger$ operators are true fermionic operators in the sense
that $\left\{a^{\dagger}_l, a_m\right\}=\delta_{l,m}$ and
$\left\{a^{\dagger}_l, a^{\dagger}_m\right\}=0=\left\{a_l, a_m\right\}$.
Furthermore, since
\begin{equation}
\s^z_l\,=\,2 a^{\dagger}_la_l -1, \label{sz}
\end{equation}
the periodic boundary conditions require that
\begin{displaymath}
\begin{array}{rclrcll}
a_{N+1} &=& - a_1,\quad & a_{N+1}^{\dagger} &=&-a_1^{\dagger} &
\quad\textrm{for states on which}\ N_f\ \textrm{is even;} \\
a_{N+1} &=& a_1, & a_{N+1}^{\dagger} &=& a_1^{\dagger} &
\quad\textrm{for states on which}\ N_f\ \textrm{is odd.}
\label{bcf}
\end{array}
\end{displaymath}

The next step is to transform the system to momentum space by performing a
Fourier transformation using new fermionic operators $b_j$ and $b^\dag_j$.
These are related to the $a_j$ and $a^\dag_j$ by
\begin{eqnarray}
b_j&=&\frac{1}{\sqrt{N}}\sum_{l=1}^N\exp\left[i k\left(j\right)l\right]\,a_l,\\
a_l&=&\frac{1}{\sqrt{N}}\sum_{j=1}^N\exp\left[-i
k\left(j\right)l\right]\,b_j, \label{vft}
\end{eqnarray}
and hermitian conjugate, where
\begin{equation}
k(j)\,=\,\frac{2 \pi}{N} \left(j\,-\,\frac{\varepsilon}{2}\right)\ \textrm{with}\,\left\{ \begin{array}{ll}
\varepsilon=0& \textrm{if}\ N_f\ \textrm{is odd},\\
\varepsilon=1& \textrm{if}\ N_f\ \textrm{is even},\quad j=1,\cdots,N.
\end{array}\right.\label{ix}
\end{equation}
The new operators $b_j$ and $b^\dag_j$ destroy or create a fermion with
momentum $k(j)$.

In terms of these operators, the Hamiltonian (\ref{vii}) is
\begin{eqnarray}
\label{x} H&=&-u\sum_{j=1}^N\cos\left[k\left(j\right)\right]\,b^\dag_j b_j{}\nonumber\\
&&{}-\frac{v}{2}\sum_{j=1}^N\cos\left[k\left(j+\frac{m}{2}\right)\right]
\left(e^{im\pi/N}\,b^\dag_j
b_{j+m}\,+\,e^{-im\pi/N}\,b^\dag_{j+m} b_{j}\right)\,-
\,h \left(\frac{N}{2}-\sum_{j=1}^N b^\dag_j b_j\right).
\end{eqnarray}
A particular case is the alternating chain that occurs when $m=N/2$.  In this
case, the Hamiltonian is
\begin{eqnarray}
H&=&-u \sum_{j=1}^N\cos\left[k\left(j\right)\right] b^\dag_j b_j{}\nonumber\\
&&{}-iv\sum_{j=1}^N\sin\left[k(j)\right]
b^\dag_j b_{j+N/2}\,- \,h \left(\frac{N}{2}-\sum_{j=1}^N
b^\dag_j b_j\right).\label{xi}
\end{eqnarray}

The commutation rule of the Hamiltonian with the $b^\dagger_j$ operator yields
\begin{eqnarray}
[H,b^\dagger_j]&=&
\left\{h-u\cos\Big[k\left(j\right)\Big]\right\}b^\dag_j\nonumber{}\\
&&{}-\,\frac{v}{2}\cos\left[k\left(j-\frac{m}{2}\right)\right]
e^{im\pi/N}b^\dag_{j-m}\,-\,
\frac{v}{2}\cos\left[k\left(j+\frac{m}{2}\right)\right]
e^{-im\pi/N}b^\dag_{j+m},
\label{xii}
\end{eqnarray}
Therefore, $b_j$ gives the hermitian conjugate of this.  As we can see, these
commutation rules mix the operators $b_j$ and $b_{j+m}$.  Hence, we can
redefine the index of $b_j$ in a new form,
\[  b_j= b_{l,\, s} \ \textrm{with}\ j = l + s\, m \ \textrm{and}\
l=1,\cdots,m \ \mbox{and}\ s=0,\cdots,a-1.
\label{xiii}
\]
The periodic conditions are $b_{l,\,a}=b_{l,\,0}$, and
$b_{l,\,-1}=b_{l,\,a-1}$.

The sums in the index $j$ will be transformed
\begin{equation}
\sum_{j=1}^N\quad\rightarrow\quad\sum_{l=1}^{m}\sum_{s=0}^{a-1}\quad
\textrm{with}\ a=\frac{N}{m},
\end{equation}
and the functions $k(j)$
\begin{equation}
k(j)\,=\,k^1(l,\,s)\,=\,k(l)+ \frac{2 \pi s}{a}
\end{equation}

Hence, the commutation rule (\ref{xii}) takes the form
\begin{eqnarray}
[H,b^\dagger_{l,\,s}]&=&
\left\{h-u\cos\left[k^1\left(l,s\right)\right]\right\}b^\dag_{l,\,s}{}\nonumber\\
&&{}-\,\frac{v}{2}\cos\left[k^1(l,s)-\frac{\pi}{a}\right]
e^{i\pi/a}\,b^\dag_{{l,\,s-1}}\,-\,
\frac{v}{2}\cos\left[k^1(l,s)+\frac{\pi}{a}\right]
e^{-i\pi/a}\,b^\dag_{{l,\,s+1}},\label{xiv}
\end{eqnarray}
that we will write as
\[
[H,\, b^\dagger_{l,\,s}]\,=\, \sum_{r=0}^{a-1} c_{r,\,s}^l
b^\dagger_{l,\,r}\label{xv}
\]
with
\begin{eqnarray}
c^l_{s,\,s}&=&h-u \cos\left[k^1(l,\,s)\right],\nonumber\\
c^l_{s-1,\,s}&=&-\frac{v}{2}\cos\left[k^1(l,\,s)-\frac{\pi}{a}\right]e^{i\pi/a},\nonumber\\
c^l_{s+1,\,s}&=&-\frac{v}{2}\cos\left[k^1(l,\,s)+\frac{\pi}{a}\right]e^{-i\pi/a},\nonumber\\
c^l_{a-1,\,0}&=&-\frac{v}{2}\cos\left[k^1(l,\,0)-\frac{\pi}{a}\right]e^{i\pi/a}, \nonumber\\
c^l_{0,\,a-1}&=&-\frac{v}{2}\cos\left[k^1(l,\,0)-\frac{\pi}{a}\right]e^{-i\pi/a},\ \textrm{and}\nonumber\\
c^l_{r,\,s}&=&0\quad\textrm{otherwise}.\label{xx}
\end{eqnarray}

For each $l$, these equations represent the elements of a hermitian $a \times
a$ matrix that we write as
\begin{equation}
C^l\,=\,\{c^l_{r,\,s}\}. \label{xxa}
\end{equation}
The eigenvalues of the these matrices yield the energy spectrum of the system.

The case $m=N/2$ does not follow from (\ref{xx}) and must be calculated
directly from the commutation relation with $a=2$.  The results for the
elements of $C^l$ are
\begin{eqnarray}
c^l_{0,\,0}&=&h-u \cos\Big[k(l)\Big],\nonumber\\
c^l_{1,\,1}&=&h+u \cos\Big[k(l)\Big],\nonumber\\
c^l_{0,\,1}&=&-i v \sin\Big[k(l)\Big].\label{xxi}
\end{eqnarray}

The Hamiltonian is diagonalized using the operators
\begin{equation}
B^\dagger_{l,\,p}\,=\,\sum_{s=0}^{a-1} s^l_{p,\,s} b^\dagger_{l,\,s}.
\label{xxii}
\end{equation}
These are defined by the solutions of the eigenvalue equation
\begin{equation}
[H,\,B^\dagger_{l,\,p}]\,=\,E^p_l\, B^\dagger_{l,\,p}. \label{xxiii}
\end{equation}
The eigenvalues $E^l_p$ are the eigenvalues of the matrix $C^l$.

Hence, the Hamiltonian becomes
\begin{equation}
H\,=\,\sum_{p=0}^{a-1}\sum_{l=1}^{m} E^p_l\, B^\dagger_{l,\,p}
B_{l,\,p} -\frac{1}{2}h N \label{xxiv}
\end{equation}

Thus, the spectrum of this Hamiltonian is grouped into $a$ bands, each of which
has $m=N/a$ levels.  In the thermodynamic limit, keeping $a$ fixed and finite,
the sums over the index $l$ become integrals.

\section{\label{results}Solutions for $\boldsymbol{a}$ = 2 and 3}

\noindent\textbf{Case} {\boldmath $a=2$}\textbf{:}

For $a=2$, the eigenvalues of $C^l$ are
\begin{eqnarray}
E_l^0&=&h-\sqrt{v^2+(u^2-v^2)\cos[k(l)]^2},\nonumber\\
E_l^1&=&h+\sqrt{v^2+(u^2-v^2)\cos[k(l)]^2}.\label{xxv}
\end{eqnarray}

\begin{figure}[t]
\includegraphics{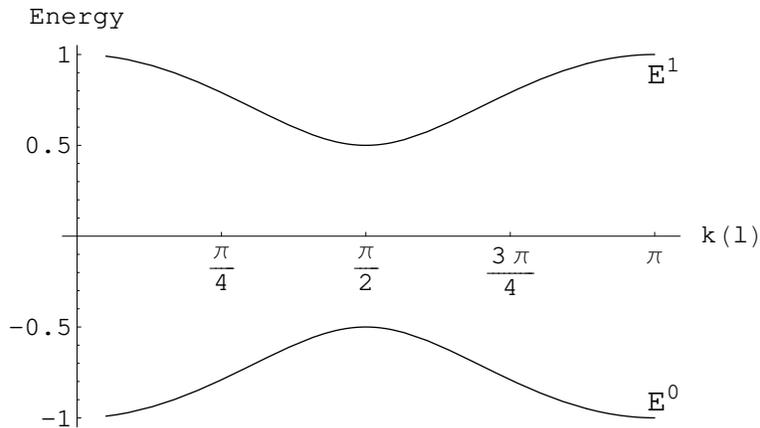}
\caption{\label{fig2}Energy of the two levels $E_0$ and $E_1$ for $u=1$ and
$v=0.5$}
\end{figure}
The parameter $l$ takes the values $1 \leq l \leq N/2$; hence, when $N
\rightarrow \infty$, $0\leq k[l]\leq \pi$.  Every value of the energy is doubly
degenerate: $E_l^i=E_{N/2-l}^i$.  In Fig.~\ref{fig2} the energy of the two
bands for the values $h=0$, $u=1$ and $v=0.5$ is shown.  For $h=0$, the lower
band has $E^0 \in [-u,\,-v]$, while the upper band has $E^1 \in [v,\,u]$.  The
difference in energy between the two bands is $2\, \min \{u,\,v\}$.  When $u=v$
both energies are independent of $l$ and every band collapses into a single
level with values $E_{0/1}\,=\,\mp u$.

\begin{figure}[b]
\includegraphics{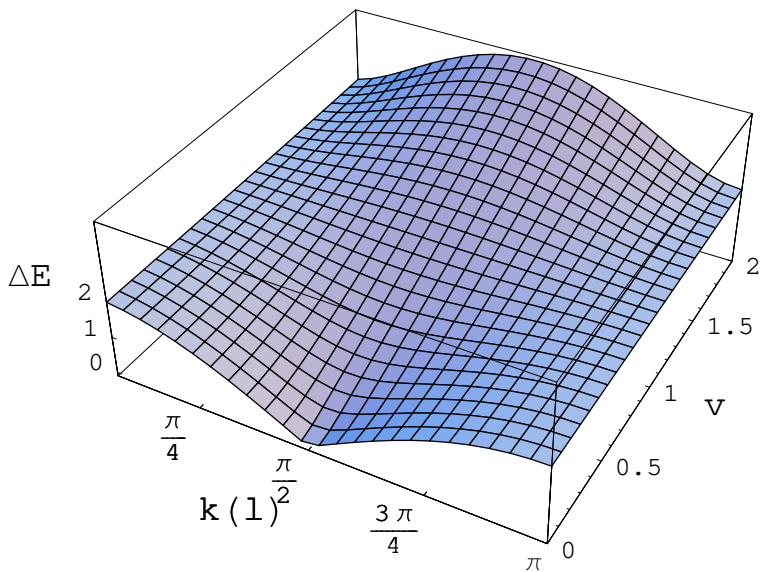}
\caption{ \label{fig1}Energy difference between the two levels
$E_0$ and $E_1$}
\end{figure}
Fig.~\ref{fig1} shows the difference,
\[
\Delta E\,=\,E_l^1\,-E_l^0\,= \,2\sqrt{u^2\cos^2k(l) +v^2 \sin^2k(l)}
\label{xxx}
\]
The the minimum of $\Delta E$ represents difference in energy between the lower
and upper bands.

\noindent\textbf{Case} {\boldmath $a=3$}\textbf{:}

In this case, the solutions for the eigenvalues of the $C^l$ are
\begin{eqnarray}
E_l^0&=& \frac{1}{48}\Big(\frac{v^{(2)}}{f^{(2)}}-f^{(2)}\Big)
  ,\nonumber\\
  E_l^1&=& \frac{1}{96}\Big((-1+i\,\sqrt{3})\frac{v^{(2)}}{f^{(2)}}+(1+i\,\sqrt{3})f^{(2)}\Big)
  \nonumber,\\
E_l^2&=&
\frac{1}{96}\Big((-1-i\,\sqrt{3})\frac{v^{(2)}}{f^{(2)}}+(1-i\,\sqrt{3})f^{(2)}\Big).
     \label{xxxii}
\end{eqnarray}
where
\begin{eqnarray}
f^{(1)}&=&\left(27648-20736\,v^2+6912\,v^3\right)\,\cos[3\,k(l)],\nonumber\\
v^{(2)}&=&-576-288\,v^2,\nonumber\\
f^{(2)}&=&\bigg(\frac{1}{2}\Big(f^{(1)}+{\sqrt{{f^{(1)}}^2 +
4\,{v^{(2)}}^3}} \Big)\bigg)^{\frac{1}{3}}.
\end{eqnarray}
These coefficients are real.

\begin{figure}[b]
\includegraphics{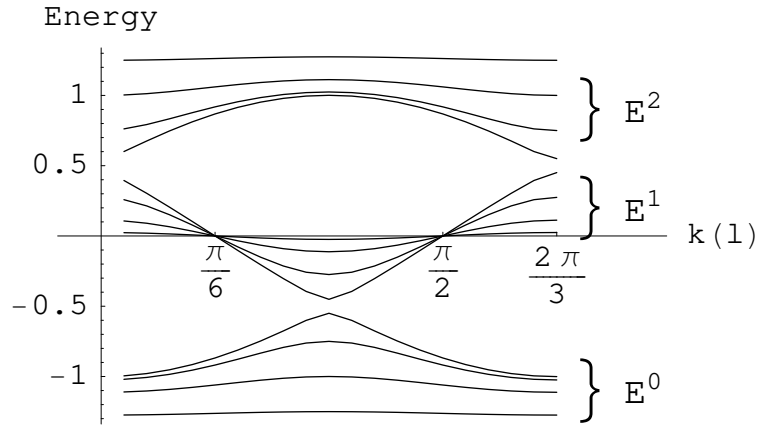}
\caption{\label{fig31}Energy of the three bands $E^0$, $E^1$ and $E^2$ for the
values $u=1$, and $v=0.1$, $0.5$, $1.0$, and $1.5$ plotted as a function of
$k(l)$.}
\end{figure}
In Fig.~\ref{fig31}, the energy levels of the three bands $E^0$, $E^1$, and
$E^2$ (\ref{xxxii}) are plotted as a function of $k(l)$ for four different
values of $v$.  The positions are marked on the $x$-axis where $k(l)$ is, for
$l$ in the interval $[1,N/3]$.

\begin{figure}
\includegraphics{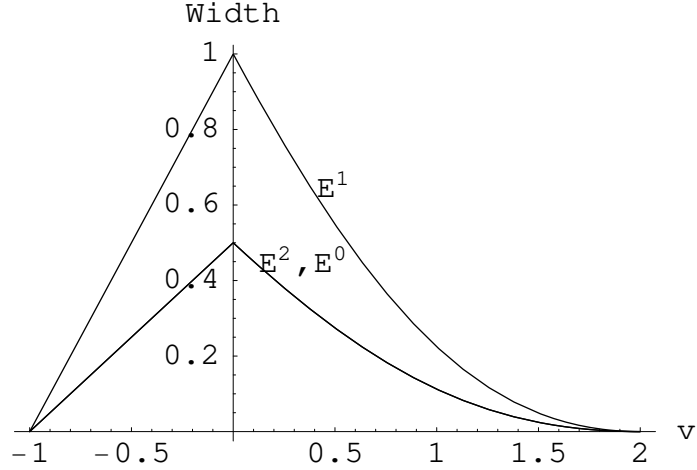}
\caption{\label{fig4rw}Energy band widths $W(E^0)=W(E^2)$ and $W(E^1)$}
\end{figure}
The width of the bands are shown in Fig.~(\ref{fig4rw}).  They are given by the
expressions
\begin{equation}
W(E^0)\,=\,\frac{W(E^1)}{2}\,=\,W(E^2)\,=\,\left\{\begin{array}{ll}
\frac{\displaystyle -1-v+\sqrt{9-6v+3v^2}}{\displaystyle 4} & \textrm{if}\ v>0\\
\frac{\displaystyle  1+v}{\displaystyle 2} & \textrm{if}\ v<0
\end{array}\right..\label{xxxiri}
\end{equation}
It can be seen from Fig.(\ref{fig4rw}) that with increasing $v$, the width of
each band increases when $v$ is negative and decreases when $v$ is positive.
The width is zero for $v=-1$ and $2$, meaning that at these two values each
band degenerates into a single level.

More generally, for an arbitrary value of $a$, the bands degenerate into single
levels when $v$ and $u$ satisfy the condition
\begin{equation}
\frac{v}{u}\,=\,-\frac{1}{\cos\left(\frac{2\pi j}{a}\right)},\quad j=1,\cdots ,a.
\label{xxxiria}
\end{equation}
When this happens, the chain splits into a number of noninteracting parts.  For
example, $j=3$ yields the solution $v=-1$, and the chain is split into
noninteracting subsystems with three sites each.  If $j=1$ or $2$, then $v=2$
and produces a chain where only the sites in the positions $3k$ and $3k+1$
($k$=integer) are interacting.

The energy difference for three bands, $\Delta
E\,=\,\min|E^2_l\,-\,E^1_{l'}|\,=\,\min|E^1_{l''}\,-\,E^0_{l'''}|$, is given by
\begin{equation}\label{{xxxirii}}
\Delta E\,=\,\left| \frac{3+3v-\sqrt{9-6v+3 v^2}}{4}\right|
\end{equation}

\begin{figure}
\includegraphics{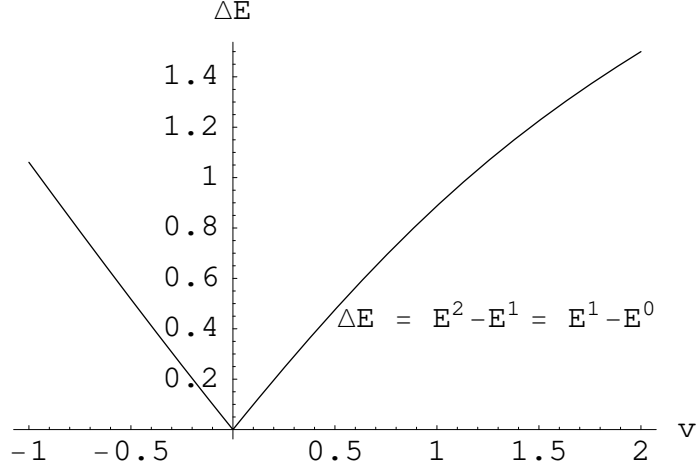}
\caption{ \label{fig32}Energy difference for three bands $\Delta E=
\min|E^1-E^0|=\min|E^2-E^1|$}
\end{figure}
$\Delta E$ over the interval $v=-1$ to $2$ is shown in Fig.~\ref{fig32}.  At
fixed $u=1$ and large $v$, $\Delta E$ is proportional to $v$:
\begin{equation}
\Delta E_{u=1,\,v\gg 2}\,=\,Gv,
\end{equation}
where the constant of
proportionality
\begin{equation}
G\,=\,\frac{3\,-\,\sqrt{3}}{4}.
\end{equation}

\pagebreak[3]\noindent\textbf{Case} {\boldmath $a>3$}\textbf{:}

\begin{figure}[b]
\includegraphics{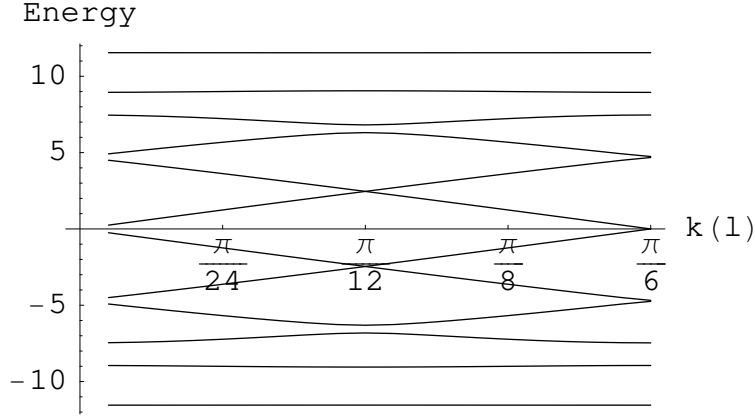}
\caption{ \label{fig12}Energy Bands for $u=10,\ v=3\ $ and $a=12$}
\end{figure}
Larger values of $a$ are straightforward to deal with.  Numerical results for
the energy bands with $a\,=\,12$, $u\,=\,10$ and $v\,=\,3$ are shown in
Fig.~\ref{fig12}.

As follows from (\ref{xxxiria}), that the bands can degenerate into a single
level only when $|v/u| \geq 1$.  If we consider only $v=\pm u$, then both cases
satisfy this condition when $a$ is even.  However, if $a$ is odd, this is true
only for $v=-u$.  In all cases that satisfy the degeneracy condition, the chain
is split into noninteracting subsystems containing $a$ sites.

\section{\label{thermo}Thermodynamic Functions}

The partition function is defined by
\begin{equation}
Z \,=\, tr\, e^{-\beta H} \,=\, \sum_{|\mathrm{est}\rangle}
e^{\,\beta hN/2}\langle\mathrm{est}|
\exp\left({\textstyle -\sum_{p,\,l}E^{p}_{l}B^{\dagger}_{l,\,p}\,B_{l,\,p}}\right)
|\mathrm{est}\rangle
\label{xxxxi}
\end{equation}
with $\beta=1/T$, and $T$ being the temperature.

The most general state of the chain can be written as a tensorial product of
the states of the sites which, in momentum space, are labelled by $\{l,\,p\}$.
The state of every site, determined by the ket $|o \rangle_{l,\,p}$, can be
occupied or empty:
\begin{equation}
|\mathrm{est}\rangle\,=\,\Pi_{l,\,p}{ |o\rangle_{l,\,p}},\qquad\textrm{with}\ o=1,\ \textrm{(occupied)}\ o=0,\ \textrm{(empty)}
\label{xxxxii}
\end{equation}

Then, the trace is,
\begin{eqnarray}
\lefteqn{\sum_{|\mathrm{est}\rangle}\langle \mathrm{est}|
\exp\left({\textstyle -\beta\sum_{p,\,l}E^{p}_{l}B^{\dagger}_{l,\,p}\,B_{l,\,p}}\right)
|\mathrm{est}\rangle}{}\nonumber\\
&&{}=\prod_{l,\,p}\left[\langle 0|
\exp\left({\textstyle -\beta E^{p}_{l}B^{\dagger}_{l,\,p}\,B_{l,\,p}}\right)
|0\rangle+\langle 1|
\exp\left({\textstyle \beta E^{p}_{l}B^{\dagger}_{l,\,p}\,B_{l,\,p}}\right)
|1\rangle\right]
\label{xxxxiii}
\end{eqnarray}

Due to the boundary conditions, the function $k(l)$ in $E_{l}^{p}$, in the
product over $l,p$, is different when the number of occupied sites $N_f$ in
$|\mathrm{est}\rangle$ is odd or even.  However, when $N\rightarrow\infty$ the
difference becomes negligible \cite{ta1,lieb}, and
\begin{equation}
Z\,=\,e^{\,\beta hN/2}\prod_{l,\,p}\left[1+\exp\left(-\beta E_{l}^{p}\right)\right].
\label{xxxxiv}
\end{equation}
This expression takes $\epsilon=0$ in $k(j)$ from (\ref{ix}) use it for
$E_{l}^{p}$.

The thermodynamic free energy per site $f$ can be derived from the partition
function as follows.
\begin{equation}
f\,=\,-\frac{1}{\beta N}\ln Z\,=\,-\frac{h}{2}-\frac{1}{\beta N}\sum_{l,\,p}\ln\left[1+\exp\left(-\beta E^p_l\right)\right].
\end{equation}

The thermodynamic limit is obtained by making $N\rightarrow\infty$ and keeping
$a$ fixed and finite, whereupon the sums become integrals:
\begin{equation}
\frac{1}{N}\sum_{l,\,p} \ \rightarrow\
\frac{1}{2\pi}\sum_{p=0}^{a-1}\int_0^{2\pi/a}\ dk.
\end{equation}

When $a=2$ the free energy can be calculated analytically.  To perform the
integrals for this case, we first change the site parameter $l$ and define
\begin{eqnarray}
E_l\,=\,E^0_l&=&\,h-\lambda(l)\quad\textrm{for}\ -\frac{N}{4}+1\leq l\leq \frac{N}{4}, \nonumber \\
E_l\,=\,E^1_l&=&\,h+\lambda(l)\quad\textrm{for}\
-\frac{N}{2}+1\leq l\leq -\frac{N}{4}\ \textrm{and}\
\frac{N}{4}+1\leq l\leq \frac{N}{2},  \label{xxxxv}
\end{eqnarray}
where
\begin{equation}
\lambda(l)\,=\,\sqrt{u^2 \cos^2 (k(l))+v^2\sin^2(k(l))}\,=\,\sqrt{v^2+(u^2-v^2)\cos^2(k(l))}.
\label{xxxxvi}
\end{equation}

Then, taking into account that $E_l$ is an even function of $l$, the free
energy in the continuum limit is
\begin{equation}
f\,=\,-\frac{h}{2}-\frac{1}{\pi \beta}\left(\int_0^{\pi/2}
\ln\left(1+e^{-\beta(h-\lambda(k))}\right)\ dk + \int_{\pi/2}^{\pi}
\ln\left(1+e^{-\beta(h+\lambda(k))}\right)\ dk \right).\label{xxxxvii}
\end{equation}

Integrating by parts yields
\begin{equation}
f\,=\,-\frac{h}{2}-\frac{1}{\beta}\left(\frac{1}{2}\ln\frac{1+e^{-\beta(h-v)}}
{1+e^{-\beta(h+v)}}+\ln\left(1+e^{-\beta(h+u)}\right)\right)
-\frac{1}{\pi}\left(I_1+I_2\right),\label{xxxxviix}
\end{equation}
where
\begin{equation}
I_1\,=\,\int^u_v \arccos\left(-\sqrt{\frac{\lambda^2-v^2}{u^2-v^2}}\,\right)
\frac{e^{-\beta(h^+\lambda)}}{1+e^{-\beta(h^+\lambda)}}\ d\lambda,
\label{xxxxviii}
\end{equation}
and
\begin{equation}
I_2\,=\,\int_{-u}^{-v} \arccos\left(\sqrt{\frac{\lambda^2-v^2}{u^2-v^2}}\,\right)
\frac{e^{-\beta(h^+\lambda)}}{1+e^{-\beta(h^+\lambda)}}\ d\lambda.
\label{xxxxix}
\end{equation}

Both integrals can be performed by standard methods: see for example the
appendix in Ref.~\cite{pat}.  Here, the function $\arccos$ is expanded as a
series and integrated term by term.  In the $I_1$ integral, we must considerer
the cases $v>-h$ and $v< -h$.  For $I_2$, the corresponding cases are $v>h$ and
$v<h$.  Then, $I_1$ in the first case is
\begin{equation}
I_1\,=\,T\left(\frac{\pi}{2}\ln\left(1+e^{-\xi}\right)+
\sum_{n=0}T^{\left(n+\frac{1}{2}\right)} a_{1,1}\left(n\right) i_{1,1}
\left(n\right)\right),\label{L}
\end{equation}
where
\begin{eqnarray}
a_{1,1}(0)&=&\frac{\sqrt{2 v}}{\sqrt{u^2 - v^2}},\nonumber\\
a_{1,1}(1)&=&\frac{3u^2+v^2}{6\sqrt{2 v (u^2-v^2)^3}},\nonumber\\
a_{1,1}(2)&=&\frac{-5u^4+50 u^2v^2+3v^4}{80\sqrt{2 v^3(u^2-v^2)^5}},\nonumber\\
a_{1,1}(3)&=&\frac{7 u^6+ 7 u^4 v^2 +301
u^2v^4+5v^6}{448\sqrt{2 v^5(u^2-v^2)^7}}, \label{Li}
\end{eqnarray}
and
\begin{equation}
i_{1,1}(n)\,=\,\int_0^{\frac{\scriptstyle u-v}{\scriptstyle T}=\infty}\frac{x^{(n+\frac{1}{2})}}{1+e^{x+\xi}}\ dx\,=\,-
\Gamma\left(\frac{3}{2} + n\right)\ \mathrm{PolyLog}\left(\frac{3}{2} + n,-e^{-\xi}\right), \label{Lii}
\end{equation}
with $\xi=(h+v)/T$.

\noindent The expression for $I_1$ when $v< -h$ is more complicated.  Following
Refs.~\cite{byr71,pru86}, this can be written as follows.
\begin{eqnarray}
I_1&=&|h|\arccos\left(-\sqrt{\frac{h^2-v^2}{u^2-v^2}}\ \right)-\frac{\pi}{2}v-u
\ \mathrm{E}\left[\arcsin\left(\frac{u}{h}\sqrt{\frac{h^2-v^2}{u^2-v^2}}\right),1-\frac{v^2}{u^2}\ \right]+\nonumber\\
&&\frac{\sqrt{(u^2-h^2)(h^2-v^2)}}{|h|}+ \sum_{n=0}
T^{n+1}\left[1-\left(-1\right)^n\right]\, a_{1,2}(n)\, i_{1,2}(n),\label{Liii}
\end{eqnarray}
where
\begin{eqnarray}
a_{1,2}(1)&=&-\frac{h}{\sqrt{(u^2-h^2)(h^2-v^2)}},\nonumber\\
a_{1,2}(3)&=&-\frac{h[2h^6-10h^2u^2v^2+h^4(u^2+v^2)+3u^2v^2(u^2+v^2)]}
{6\sqrt{(u^2-h^2)^5(h^2-v^2)^5}},\label{Liiii}
\end{eqnarray}
and
\begin{equation}
i_{1,2}(n)\,=\,\int_0^{\infty}\frac{x^n}{1+e^x}\ dx
\,=\,\frac{2^n-1}{2^n}\ \Gamma(n+1)\ \zeta(n+1).\label{Liiiir}
\end{equation}
The first three elements of~(\ref{Liiiir}) are
\begin{equation}
i_{1,2}(1)\,=\,\frac{\pi^2}{12},\qquad
i_{1,2}(3)\,=\,\frac{7\pi^4}{120},\qquad
i_{1,2}(5)\,=\,\frac{31\pi^6}{252}.\label{Lv}
\end{equation}
Once again, $|u-h|/T$ and $|h-v|/T$ are assumed to be large.

The $I_2$ integral when $h<v$ can also be performed using the tables provided
in Refs.~\cite{byr71,pru86}, and a similar expansion to before.  The result is
\begin{eqnarray}
I_2&=&(u-v)+u\ \mathrm{E}\left(\sqrt{1-\frac{v^2}{u^2}}\ \right)-T
\frac{\pi}{2}\ln\left(1+e^{\left(h-v\right)/T}\right)\nonumber\\
&&-T\sum_{n=0}T^{n+\frac{1}{2}} a_{2,1}\left(n\right)
i_{2,1}\left(n\right),\label{Lvi}
\end{eqnarray}
where $a_{2,1}(n)=-a_{1,1}(n)$ from (\ref{Li}), and $i_{2,1}(n)=i_{1,1}(n)$
from (\ref{Lii}), with $\xi=(v-h)/T$.

The result for integral $I_2$ when $h>v$ is
\begin{eqnarray}
I_2&=&h\arccos\left(\sqrt{\frac{h^2-v^2}{u^2-v^2}}\ \right)+u
\ \mathrm{E}\left[\arccos\left(\sqrt{\frac{h^2-v^2}{u^2-v^2}}\right),
\sqrt{1-\frac{v^2}{u^2}}\ \right]\nonumber\\
&&+\sum_{n=0} T^{n+1}\left[1-\left(-1\right)^n\right]\,
a_{2,2}(n)\, i_{2,2}(n),\nonumber\label{Lvii}
\end{eqnarray}
where $a_{2,2}(n)=-a_{1,2}(n)$ from (\ref{Liiii}), and $i_{2,2}=i_{1,2}$ from
(\ref{Liiiir}).

\begin{figure}
\includegraphics[scale=.8]{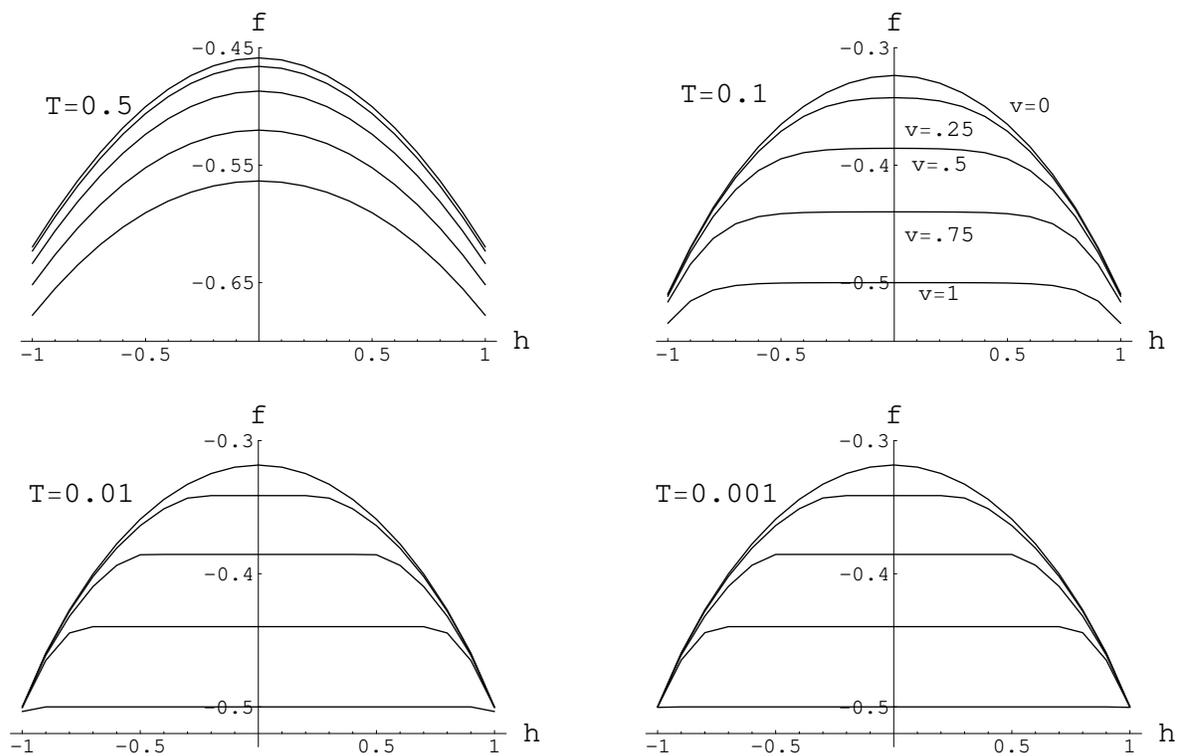}
\caption{\label{fig41}Free Energy plotted using $u=1$ and $v=0.0$, $0.25$,
$0.5$, $0.75$, $1$.}
\end{figure}
The free energy is obtained by substituting $I_1$ and $I_2$ into
(\ref{xxxxviix}).  Several examples are plotted in Fig.~\ref{fig41} for
different values of $T$ and $v$, and for $u=1$, as functions of $h$.  These
show that, as implied by the Hamiltonian (\ref{hxx}), the free energy is an
even function of $h$.  This property is also shown explicitly by an alternative
expression to (\ref{xxxxvii}) for the free energy:
\begin{equation}
f\,=\,-\frac{T}{\pi}\int_0^{\frac{\pi}{2}}\ln\left(2\cosh\frac{h}{T}+2\cosh\frac{\lambda(k)}{T}\right)\ dk.\label{Lviir}
\end{equation}
Notice that at low $T$, $f$ becomes nearly constant between $-v$
and $v$.

The entropy per site is defined by
\begin{equation}
s\,=\,-\frac{\partial f}{\partial T}.\label{Lviii}
\end{equation}
This can be obtained exactly from the expression for the free energy.
To examine its behaviour we evaluate the partial derivative as follows.
\begin{equation}
s\,=\,\frac{1}{\pi}\left[
\int_0^{\frac{\pi}{2}}w\left(\frac{h-\lambda(x)}{T}\right)\ dx\,+\,
\int_{\frac{\pi}{2}}^{\pi}w\left(\frac{h+\lambda(x)}{T}\right)\ dx\right],
\label{Lix}
\end{equation}
where the function $w(x)$ is
\begin{equation}
w(x)\,=\,\ln\left(1+e^{-x}\right)+\frac{x e^{-x}}{1+ e^{-x}}.\label{Lx}
\end{equation}
This is an even function.  Its magnitude is significant only when $x$ is small:
$(w(5)=0.04$, $w(10)=0.0004)$.  For large values of $|x|$, the expression
simplifies to
\begin{equation}
w(x)\,=\,{|x| e^{-|x|}}.\label{Lxr}
\end{equation}

The integral can be evaluated in a similar manner to the one for the free
energy.  However, in this case the properties of the function $w$ allow us to
examine the behaviour of $s$ in the low temperature limit.  When $|h|< v$, the
integrand in (\ref{Lix}) can be approximated by (\ref{Lxr}).  Taking the most
significant term, yields the following solution.

\begin{equation}
s\left(h,T\right)_{T\rightarrow 0}\,\approx\,\frac{1}{\pi}\,
e^{-\left(v-|h|\right)/T}\,\rightarrow\,0.\label{Lxr1}
\end{equation}
A similar approximation can be made for $|h|>u$:
\begin{equation}
s\left(h,T\right)_{T\rightarrow 0}\,\approx\,\frac{1}{\pi}\,
e^{-\left(|h|-u\right)/T}\,\rightarrow\,0.\label{Lxrr1}
\end{equation}
To evaluate (\ref{Lix}) for other values of $h$, we change the integration
variable to
\begin{equation}
y\,=\,\frac{h-\lambda(x)}{T},\quad\textrm{and}\quad
y\,=\,\frac{h+\lambda(x)}{T}\label{Lxi}
\end{equation}
respectively, in the two integrals.  Hence, the entropy is
\begin{eqnarray}
s&=&\frac{1}{\pi}\Bigg(\int_{(h-u)/T}^{(h-v)/T}
w(y)\ \frac{T(h-Ty)}{\sqrt{[u^2-(h-Ty)^2][(h-Ty)^2-v^2]}}\ dy{}\nonumber\\
&&{}+\,\int_{(h+v)/T}^{(h+u)/T}w(y)\
\frac{T(Ty-h)}{\sqrt{[u^2-(h-T y)^2][(h-T y)^2-v^2]}}\
dy\Bigg).\label{Lxii}
\end{eqnarray}
This integral has local maxima at $|h|=v$, where it can be approximated by
\begin{equation}
s(|h|=v)\,=\,\frac{1}{\pi}\,\sqrt{\frac{T v}{2(u^2-v^2)}}\,
\int^{\infty}_{0}\,\frac{w(y)}{\sqrt{y}}\ dy\,=\,0.6475\,
\sqrt{\frac{T v}{2(u^2-v^2)}}.\label{Lxiir}
\end{equation}
A similar approximation can be made when $|h|=u$, where there is also a local
maximum.
\begin{equation}
s(|h|=u)\,=\,0.6475\,\sqrt{\frac{Tu}{2(u^2-v^2)}}\label{Lxiirr}
\end{equation}

When $v<|h|<u$ ---taking into account that $w(x)$ has significant
values around $x \approx 0$, and is rapidly decreasing
otherwise--- $s$ can be approximated by
\begin{equation}
s\,=\,\frac{1}{\pi}\frac{T|h|}{\sqrt{(u^2-h^2)((h^2-v^2)}}
\int_{-\infty}^{\infty}w(y)\ dy
\,=\,\frac{\pi T |h|}{3\sqrt{(u^2-h^2)(h^2-v^2)}}.\label{Lxiii}
\end{equation}
Here, we can use
\begin{equation}
\frac{1}{\pi}\int_{0}^{\infty}\frac{w(x)}{\sqrt{x}}\ dx
\,=\,0.6475,\qquad\frac{1}{\pi}\int_{-\infty}^{\infty}w(x)\ dx
\,=\,\frac{\pi}{3}.
\end{equation}
Thus, combining all of the above, the entropy is
{\setlength\arraycolsep{1em}
\begin{equation}
s(h,T)_{T\rightarrow 0}\,=\,\left\{
\begin{array}{ll}
\,{\frac{\displaystyle 1}{\displaystyle \pi}\,
e^{-\left(v-|h|\right)/T}}\quad\rightarrow\quad 0
&\textrm{if}\ |h|<v\\
0.6475\ \sqrt{\frac{\displaystyle Tv}{\displaystyle 2(u^2-v^2)}}
&\textrm{if}\ |h|=v\\
\frac{\displaystyle \pi T|h|}{\displaystyle 3\sqrt{(u^2-h^2)(h^2-v^2)}}
&\textrm{if}\ u>|h|>v\\
0.6475\ \sqrt{\frac{\displaystyle Tu}{\displaystyle 2(u^2-v^2)}}
&\textrm{if}\ |h|=u\\
{\frac{\displaystyle 1}{\displaystyle \pi}\,
e^{-\left(|h|-u\right)/T}}\quad\rightarrow\quad 0
&\textrm{if}\ |h|>u
\end{array}\right..\label{Lxiv}
\end{equation}}

\begin{figure}[b]
\includegraphics[scale=1]{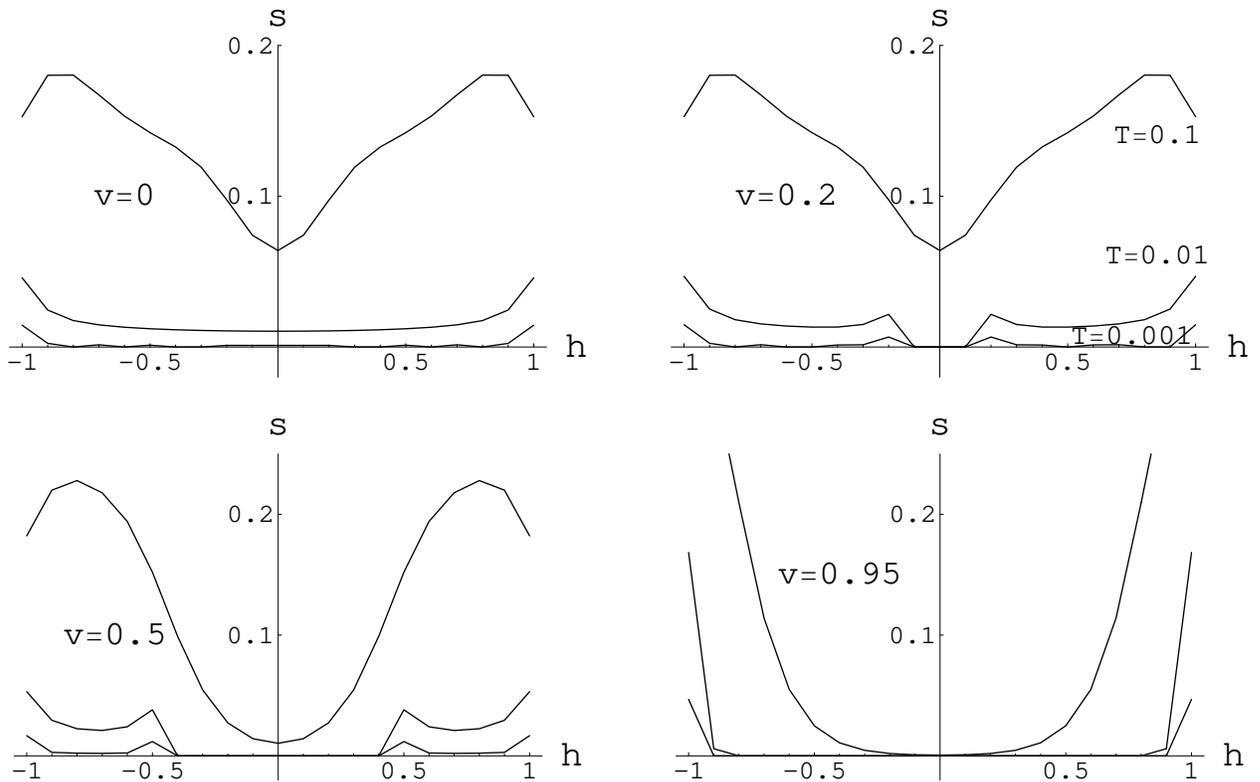}
\caption{\label{fig42}Entropy plotted using $u=1$ and $T=0.1$,
$0.01$, $0.001$. The ordering of the lines for $v=0.2$ is the same
in the other three cases.}
\end{figure}
The entropy is plotted using four different values of $v$, each at three
different temperatures in Fig.~\ref{fig42}.
\begin{figure}[t]
\includegraphics{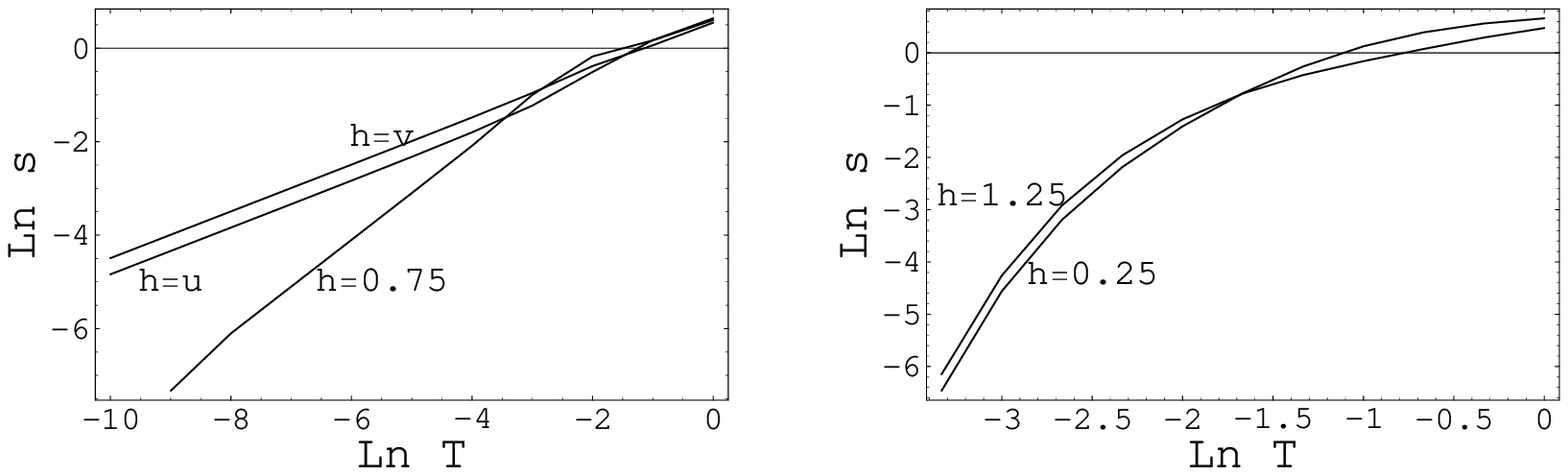}
\caption{\label{fig9}Graphs of $\ln S$ versus $\ln T$ plotted using $u=1$ and
$v=0.5$, for five values of $h$.  Left: $h=1$, $0.75$, $0.5$; and right:
$h=1.25$, $0.25$.}
\end{figure}
Fig.~\ref{fig9} shows logarithmic graphs of the temperature dependence of the
entropy when $u=1$ and $v=0.5$, for five values of values $h$.

The specific heat per site is
{\setlength\arraycolsep{1em}
\begin{equation}
C(h,T)_{T\rightarrow 0}\,=\,T\frac{\partial s}{\partial T}\,=\,\left\{
\begin{array}{ll}
\,(\frac{\displaystyle v-|h|}{\displaystyle T})\,e^{-\left(v-|h|\right)/T}
&\textrm{if}\ |h|<v\\
0.3238\ \sqrt{\frac{\displaystyle Tv}{\displaystyle2(u^2-v^2)}}
&\textrm{if}\ |h|=v\\
\frac{\displaystyle\pi T|h|}{\displaystyle 3\sqrt{(u^2-h^2)(h^2-v^2)}}
&\textrm{if}\ |h|>v\\
0.3238\ \sqrt{\frac{\displaystyle Tu}{\displaystyle 2(u^2-v^2)}}
&\textrm{if}\ |h|=v\\
\frac{\displaystyle |h|-u}{\displaystyle T}\,e^{-\left(|h|-u\right)/T}
&\textrm{if}\ |h|>u
\end{array}\right..\label{Lxv}
\end{equation}}

It can be seen that both functions (the entropy and the specific heat) decay
exponentally as $e^{-1/T}$ to zero when $T\rightarrow 0$, for $|h|<v$ and
$|h|>u$.
 This also shows that the
internal parameters of the system, $v$ and $u$, can be derived by measuring the
specific heat as a function of $h$.

Conclusions for other values of periodicity $a$ can be generalised from this
result.  The specific heat will be exponentially small for values of $h$
between the several energy bands.  Thus, measurements will provide information
about the periodicity and its amplitude.

\section*{Acknowledgments}

We would like to thank Dr.\
A.~Cruz for his comments.  This work has been supported by the
CICYT grants BFM2000-1057 and FPA2000-1252.

\bibliography{XXref}
\bibliographystyle{unsrt}
\end{document}